# Simulation of stripping injection into HITFiL with carbon ion

XIE Xiu-Cui(谢修璀)[1,2]   SONG Ming-Tao(宋明涛)[1]   ZHANG Xiao-Hu(张小虎)[1,2]

1 Institute of Modern Physics, Chinese Academy of Sciences, Lanzhou 730000, China

2 University of Chinese Academy of Sciences, Beijing 100049, China

**Abstract:** Stripping injection is one of the crucial stages in the accumulation process of the hadron therapy synchrotron HITFiL (Heavy Ion Therapy Facility in Lanzhou). In order to simulate the stripping injection process of carbon ions for HITFiL, the interactions between carbon ions and foil has been studied, and simulated with a code developed by ourselves .The optimized parameters of the injecting beam and the scheme of the injection system have been presented for HITFiL.

## 1 Introduction

Hadron therapy with carbon ions, with its Bragg curve dose deposit which is a significant advantage compared with the conventional radiotherapy methods, makes it one of the most popular and important new developments in today's medical research and accelerator technology [1]. In developed countries such as USA, Germany and Japan, hadron therapy has finished its experimental stage and is becoming a more and more widely-used medical method [2].

HITFiL, the Heavy Ion Therapy Facility in Lanzhou, is a compact hadron therapy facility consisting of a cyclotron injector and a 56.173m synchrotron. Due to its special attention to low cost and high reliability, the stripping injection, which has relative lower cost as well as less complexity compared with multiturn injection with electron cooling, becomes the best compromise choice [3].

The carbon ions $C^{5+}$ of 7MeV/u, extracted from the cyclotron injector, are transported to the synchrotron by a middle energy beam transport line. A carbon stripping foil is placed at an angle of 12 degrees behind the entrance of the first dipole magnet, as shown in Fig.1.

The intensity of the injected carbon ions is 10 μA, which means there are $2\times10^7$ particles injected into the ring every turn. The prospective stored particles of the ring are $1\times10^9$, which could be achieved by 50 times accumulation.

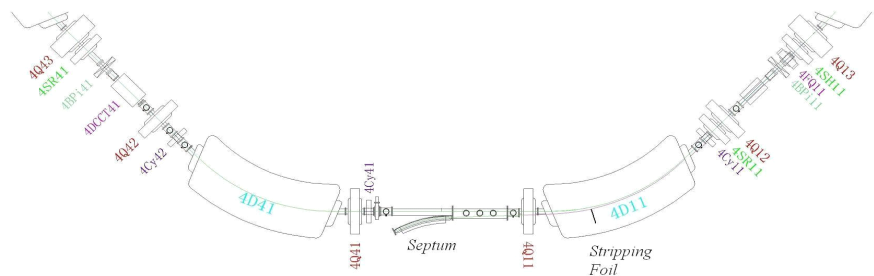

**Fig.1. Layout of the injection section of HITFiL**

## 2 Interaction between injected ions and stripping foil

When the injected ions pass through the stripping foil, several processes take place, mainly the inelastic collisions with the atomic electrons of the foil and the elastic scatterings from nuclei, which can result in much energy loss, change of ion charge state and increase of beam emittance. Other processes such as emission of Cherenkov radiation, nuclear reactions or bremsstrahlung can also occur. But they are negligible [4] within the energy range of our machine.

### 2.1 Stripping efficiency

When a beam of particles pass through a foil, electrons of the moving particles can be stripped away or the particles can capture electrons from the foil atoms [5]. The fraction $F_i$ of a charge state i can be expressed by:

$$\frac{dF_i}{dx} = \sum_{j \neq i} \sigma_{ji} F_j - \sum_{j \neq i} \sigma_{ij} F_i, \quad (1)$$

where $\sigma_{ij}$ is the cross section from the charge states i to j and x is the foil thickness with unit of particle number per square meter.

When the velocity of an injected particle is much larger than the Thomas-Fermi speed, the capture cross section is much smaller than the stripping cross section and can be neglected. So, we have the following stripping cross section:

$$\sigma_s = 8\pi a_0^2 \left(\alpha/\beta\right) S, \quad (2)$$

where $\alpha \approx 1/137$ is the fine structure constant, $a_0 = 5.29 \times 10^{-11}$ m is the Bohr radius, and $\beta$ is the relativistic factor. Moreover,

$$S \approx \frac{1.24 Z_T}{Z_p^2} \left(1 + 0.105 Z_T - 5.4 \times 10^{-4} Z_T^2\right) \quad (3)$$

is the collision strength introduced to fit the experimental data, where $Z_T$ is the electron number of the foil atoms and $Z_P$ is the charge state of the projectiles.

So, we can work out the relationship between the stripping efficiency and the foil thickness shown in Fig.2.

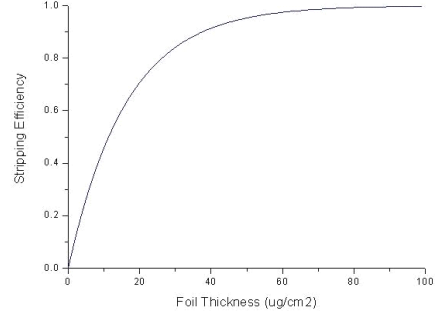

**Fig.2. Stripping efficiency vs foil thickness.**

In particular, we see that a 20μg/cm² foil has the stripping efficiency 70.7% and a 30μg/cm² foil has the efficiency 84.1%.

### 2.2 Energy loss and straggling

In the process of inelastic collisions, energy is transferred from the injected particles to the foil atoms, this means the average energy loss of a particle is governed by the famous Bethe-Block formula:

$$\langle E \rangle = \xi \left[ Ln\left(\frac{2 m_e c^2 \beta^2 \gamma^2 E_{max}}{I^2}\right) - 2\beta^2 \right], \quad (4)$$

where

$$\xi = \frac{0.30058 \cdot Z_T m_e c^2 t}{\beta^2 A_T} \text{ MeV}. \quad (5)$$

In the above expression, $m_e c^2$ is the electron mass expressed in MeV, t is the foil thickness in g/cm², and $A_T$ is the atomic weight of the foil material [6]. $E_{max}$ is the maximum energy transferable to an atomic electron, given by:

$$E_{max} = 2 m_e c^2 \beta^2 \gamma^2 \left[1 + 2\gamma \frac{m_e}{m_p} + \left(\frac{m_e}{m_p}\right)^2\right]^{-1}, \quad (6)$$

where $m_p$ is the proton rest energy, γ is the relativistic factor and I is the mean excitation potential of the absorber atom.

Due to its Landau distribution, an energy loss can be generated by the formula [6]:

$$\delta E = \langle E \rangle + \xi [DINLAN(R) + 1 + \beta^2 - 0.577216 + Ln\kappa], \quad (7)$$

where $\kappa = \xi / E_{max}$ is the energy loss parameter and DINLAN is a CERN routine which computes the inverse of Landau cumulative distribution function and is controlled by a random number R (0≤R≤1)

Then, we have the relation between energy straggling and foil thickness shown in Fig.3(a).

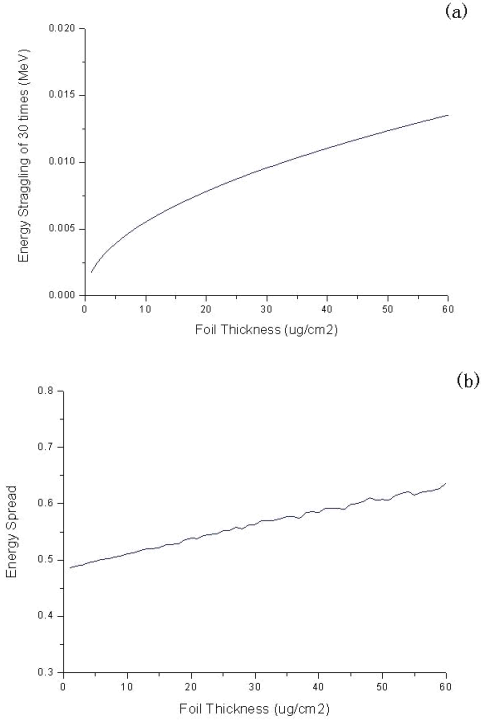

**Fig.3. (a) Energy straggling vs foil thickness; (b) Energy spread vs foil thickness.**

Because the energy straggling in one particle could bring the whole beam energy spread, a Monte-Carlo simulation is performed with 100,000 macro-particles. The result is shown in Fig.3(b).

### 2.3 Angular scattering

The Coulomb elastic scattering causes a change in the ion's moving direction, the mean square angle of the multiple scatterings can be described as the following experimental formula:

$$\langle \Delta \theta^2 \rangle = 0.25 \frac{Z_T (Z_T + 1)}{A_T} \frac{Z_P^2}{E_P^2} t, \quad (8)$$

where the scattering angle θ is in rad and the projectile energy $E_P$ is in MeV [7].

This leads to the results shown in the following pictures.

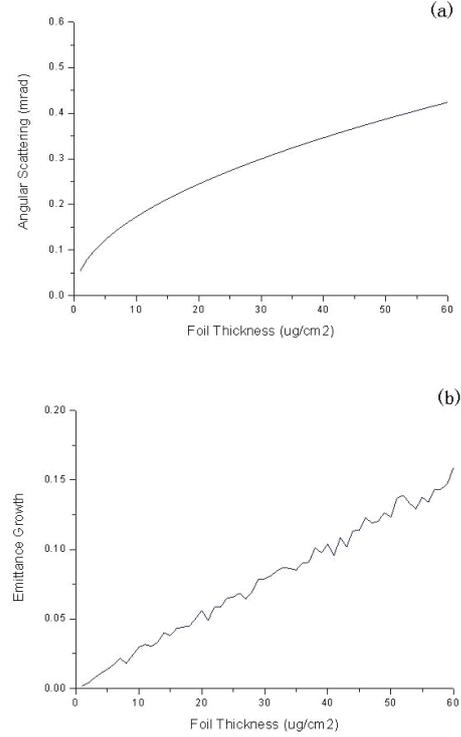

**Fig.4. (a) Angular scattering vs foil thickness; (b) Emittance growth vs foil thickness.**

The upper picture shows the relationship between the foil thickness and the angular scattering, the lower one shows the emittance growth in the transversal direction caused by the angular scatterings which are simulated by a 100,000-particle with the Monte-Carlo method.

There are still some other effects such as lateral spread, which are relatively small in quantity and can be ignored.

### 3 The injection simulation

With the above study of particle-foil interaction, a simulation code is developed using Matlab. The injected $C^{5+}$ particles with certain distribution or Twiss parameters are produced by the Mont-Carlo method. A checking subroutine is activated to check whether these particles will hit the foil and will be stripped to $C^{6+}$ or not. The circulating particles, together with the newly stripped ones (the particles injected and stripped in the first turn do not have circulating particles existing), will be added with their energy losses, angular scatterings and lateral spreads as they pass through the foil. With all these particles, their coordinates in every direction (X, Xp, Y, Yp, dE and phase) will be written into a MAD [8] input file where the synchrotron model has already been constructed to start a tracking subroutine. The coordinates output from the tracking will be written back to start a new turn of tracking until the whole injection process finishes. To verify its reliability, the code has been modified to proton injection simulation and checked with ACCSIM [9] to prove its accuracy.

In order to make the accumulation process more clear, the interface has been made with ACCSIM style which is shown in Fig.5.

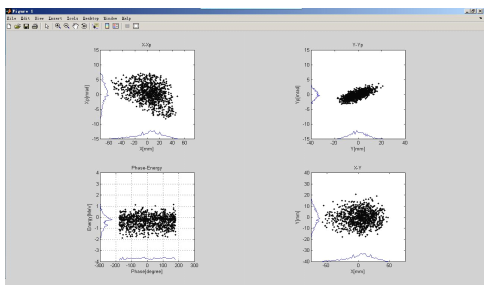

Fig.5. Interface of the simulation code.

### 3.1 Foil thickness

With the help of the simulation code, we firstly try to work out how the foil thickness affects the injection efficiency. A thick foil usually has a higher stripping efficiency which is good for the accumulation but causes more energy loss which is bad for the accumulation, while a thin one has the opposite effect. During the simulation, we found that a carbon stripper with the thickness of 20-25 $\mu g/cm^2$ is the best choice. A comparison of particle accumulations among different foil thicknesses is shown in Fig.6.

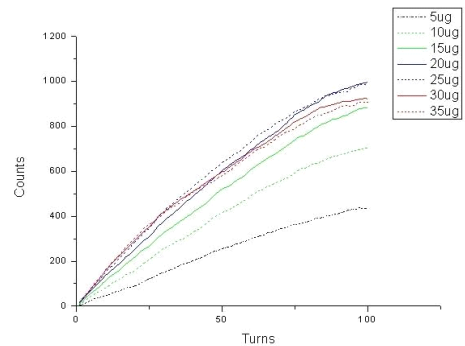

Fig.6. Particle accumulation of different foil thicknesses. The macro-particles injected in every turn are 20.

### 3.2 Bump curve

A different bump curve usually has a different impact on the accumulation. We choose three typically different curves to examine their impacts: a linear bump, a cosine bump and an exponential bump. A comparison among them is shown in Fig.7 where we can see the linear bump curve and cosine bump curve almost have the same accumulation efficiency while the exponential one has a lower efficiency.

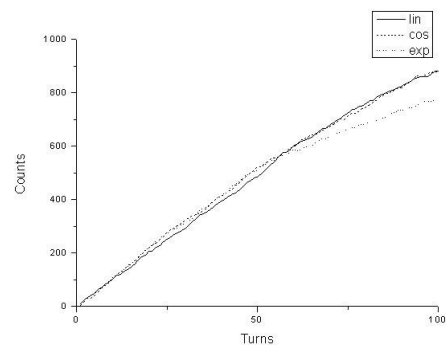

Fig.7. Particle accumulation of different bump curves. The macro-particles injected in every turn are 20.

### 3.3 Beam quality

The momentum spread of the injecting beam, which is defined as its quality, plays a quite important role in the injecting process. In order to find its effect on the accumulation, we compare four sets of different momentum spread: the 0.1%, 0.25%, 0.5% momentum spread beams with design central energy (84MeV) and the 0.1% momentum spread beam with 0.5MeV higher than the designed central energy (84.5MeV). The result is shown in Fig.8, where the 0.1% momentum spread beam with 0.5MeV higher energy has the best accumulation efficiency. This is because they lose some of their energy and slowly approach the designed energy when the particles with higher energy hit the foil.

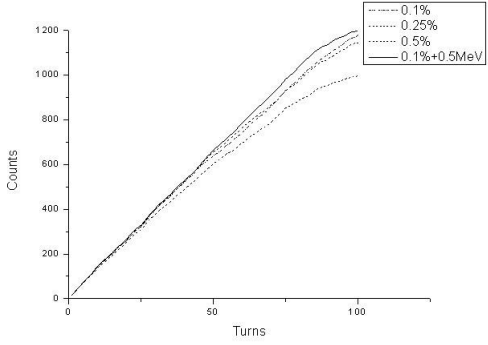

**Fig.8. Particle accumulation of different momentum spread. Every turn 20 macro-particles are injected into the ring**

### 3.4 Injection matching

The matching of Twiss parameters is a very important factor in the injection. Generally we have two principles:

$$\frac{\alpha_i}{\beta_i} = \frac{\alpha_m}{\beta_m} = -\frac{X'_c - X'_o}{X_c - X_o}, \quad (9)$$

$$\frac{\beta_i}{\beta_m} \geq \left(\frac{\varepsilon_i}{\varepsilon_m}\right)^{\frac{1}{3}}, \quad (10)$$

where $\alpha_i$, $\beta_i$, $\varepsilon_i$ are the Twiss parameters and the emittance of the injecting beam while $\alpha_m$, $\beta_m$, $\varepsilon_m$ are the Twiss parameters and the emittance of the circulating beam. And the subscripts C and O indicate the injection point and the close orbit respectively [10].

In order to find the best set of matching parameters and determine whether they are a steady one, a scan of the Twiss parameters together with their accumulation efficiencies is presented in Fig.9. We can see that there is a flat slope on the top of the "mountain" which indicates the steady region of injecting Twiss parameter.

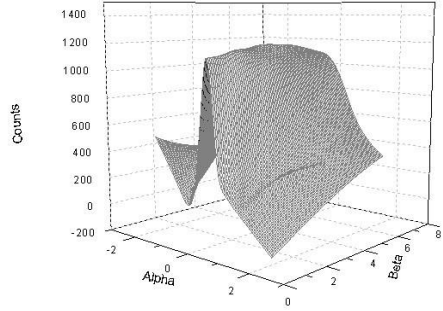

**Fig.9. Particle accumulation of different matching parameters. 20 macro-particles are injected into the synchrotron every turn.**

### 4 Conclusion

In this paper, we present the interaction between the injected carbon ions $C^{5+}$ and the stripping carbon foil. Also, a simulation code is developed to study the impact of the injecting beam parameters as well as the injection scheme setting. These results would be very important and helpful for the optimization study and commissioning work of HITFiL in the near future.